\documentclass[12pt,preprint]{aastex}

\usepackage{epsfig}

\begin{document}
\title{Initial Planetesimal Sizes and the Size Distribution of Small Kuiper Belt Objects}
\shortauthors{Schlichting et al.} 
\shorttitle{The Size Distribution of Small Kuiper Belt Objects}
\author{Hilke E. Schlichting\altaffilmark{1,2,3}, Cesar I. Fuentes\altaffilmark{4}, David E. Trilling\altaffilmark{4}}

\altaffiltext{1} {UCLA, Department of Earth and Space
Science, 595 Charles E. Young Drive East, Los Angeles, CA
90095, USA}\altaffiltext{2} {Department of Astronomy, California Institute of Technology, MC 130-33,
Pasadena, CA 91125, USA} 
\altaffiltext{3} {Hubble Fellow}
\altaffiltext{4}{Department of Physics and Astronomy, Northern Arizona University, P.O. Box 6010, Flagstaff, AZ 86011, USA}
\email{hilke@ucla.edu}

\begin{abstract} 
The Kuiper Belt is a remnant from the early solar system and its size distribution contains many important constraints that can be used to test models of planet formation and collisional evolution. We show, by comparing observations with theoretical models, that the observed Kuiper Belt size distribution is well matched by coagulation models, which start from an initial planetesimal population with radii of about 1~km, and subsequent collisional evolution. We find that the observed size distribution above $R \sim 30$~km is primordial, i.e., it has not been modified by collisional evolution over the age of the solar system, and that the size distribution below $R \sim 30$~km has been modified by collisions and that its slope is well matched by collisional evolution models that use published strength laws. We investigate in detail the resulting size distribution of bodies ranging from 0.01~km to 30~km and find that its slope changes several times as a function of radius before approaching the expected value for an equilibrium collisional cascade of material strength dominated bodies for $R \lesssim 0.1$~km. Compared to a single power law size distribution that would span the whole range from 0.01~km to 30~km, we find in general a strong deficit of bodies around $R\sim 10$~km and a strong excess of bodies around 2~km in radius. This deficit and excess of bodies are caused by the planetesimal size distribution left over from the runaway growth phase, which left most of the initial mass in small planetesimals, while only a small fraction of the total mass is converted into large protoplanets. This excess mass in small planetesimals leaves a permanent signature in the size distribution of small bodies that is not erased after 4.5 Gyrs of collisional evolution. Observations of the small KBO size distribution can therefore test if large KBOs grew as a result of runaway growth and constrain the initial planetesimal sizes. We find that results from recent KBO occultation surveys and the observed KBO size distribution can be best matched by an initial planetesimal population that contained about equal mass per logarithmic mass bin in bodies ranging from 0.4~km to 4~km in radius. We further find that we cannot match the observed KBO size distribution if most of the planetesimal mass was contained in bodies that were 10~km in radius or larger, simply because their resulting size distribution cannot be sufficiently depleted over 4.5 Gyrs to match observations. 
\end{abstract}


\section{INTRODUCTION}
The Kuiper Belt consists of a disk of icy objects located just beyond the orbit of Neptune. In the Kuiper Belt planet formation never proceeded all the way to completion, which makes it an ideal laboratory for testing planet formation theories. 

\begin{figure} [htp]
\centerline{\epsfig{file=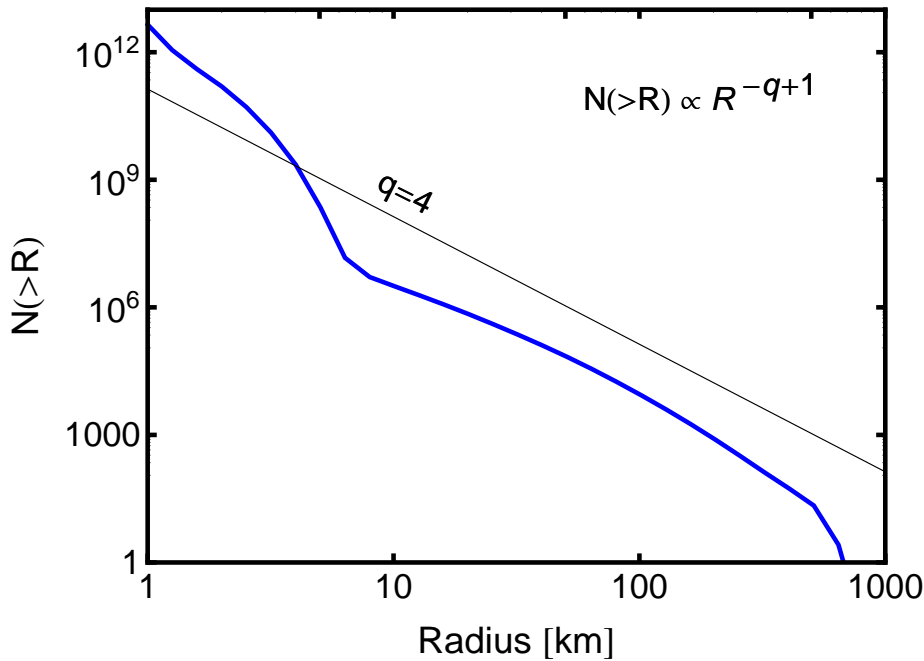, scale=1.2}}
\caption{The size distribution at the end of runaway growth before the onset of collisional erosion is given by the thick blue line. Note that during runaway growth, most of the initial mass remains in small planetesimals, while a small fraction of the total mass is converted into large protoplanets/KBOs. This specific example corresponds to an initial planetesimal population of bodies that were all 1~km in radius. As shown in Figures \ref{fig1} and \ref{fig2}, the current large KBO size distribution is well matched by the resulting size distribution from runaway growth. For comparison, a power-law size distribution with differential power-law index $q=4$ is given by the thin black line.}
\label{fig4}
\end{figure}

The Kuiper belt size distribution contains many important clues
concerning the formation of Kuiper belt objects (KBOs), their effective
strength, and their collisional evolution \citep{D69,DF97,KL99,PS05}. The cumulative size distribution of KBOs larger than $R
\gtrsim 30~\rm{km}$ (i.e., objects with R-band magnitudes brighter than about 25) is well described by a single power-law given by
\begin{equation}\label{e2}
N(>R) \propto R^{1-q}
\end{equation} 
where $N(>R)$ is the number of objects with radii greater than $R$, and $q$ is
the power-law index. Kuiper belt surveys find that the size distribution for
KBOs with radii greater than about 30~km follows this power-law with $q \sim
4$ \citep[e.g.][]{TJL01,BTA04,FH108,FK108}, which implies roughly equal mass
per logarithmic mass interval. This size distribution is a relic of the
accretion history in the Kuiper belt and therefore provides valuable insights
into the formation of large KBOs ($R \gtrsim 30~\rm{km}$) \citep[e.g.][]{S96,DF97,KB04,SS11}. It has been shown by several works that the large KBO size distribution can be well matched by numerical coagulation simulations \citep[e.g.][]{KL99,SS11,KB12}. For example, \citet{SS11} find that the size distribution of larger KBOs is well matched by planet formation models of runaway growth. During runaway growth only a small fraction of the total mass is converted into large protoplanets, while most of the initial mass remains in small
planetesimals. The size distribution of the large protoplanets in the runaway tail follows a power  law size distribution with differential power law index $q\sim4$, implying roughly equal mass per logarithmic mass bin (see Figure \ref{fig4}). 

\begin{figure} [htp]
\centerline{\epsfig{file=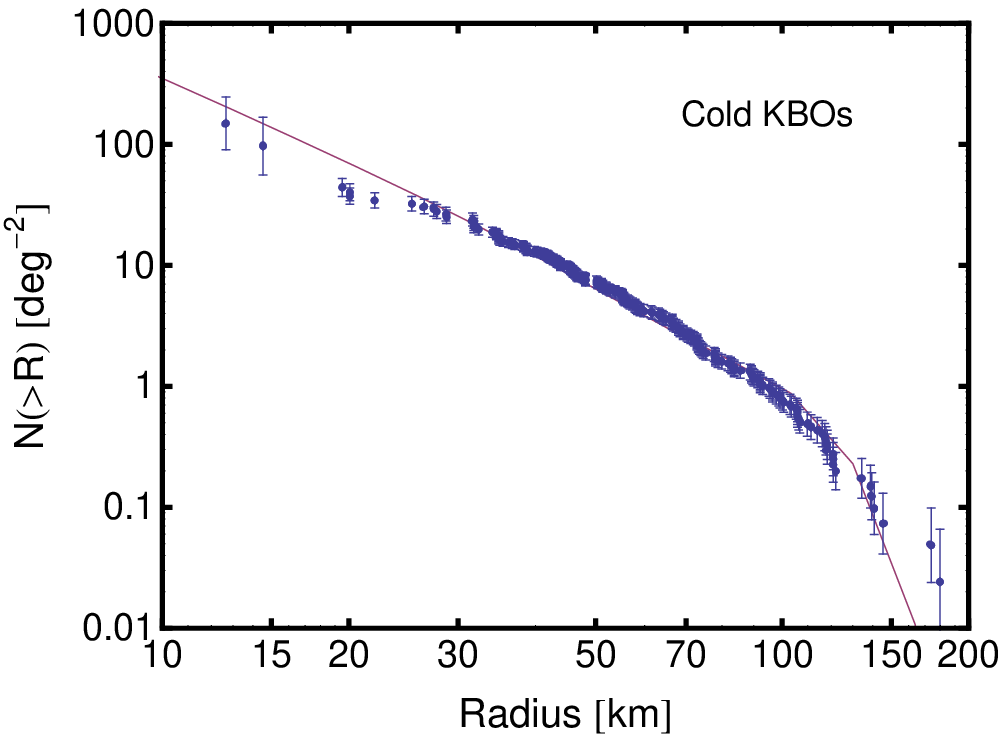, scale=1.2}}
\caption{Comparison between the observed Kuiper Belt size distribution for objects with inclinations $\leq 5^{\circ} $, also referred to as the cold population, as summarized in \citet{FH10} (points), with the numerical coagulation results from \citet{SS11} (line). The error bars give the $1 \sigma$ errors on the cumulative size distribution. The observed Kuiper Belt size distribution above $R \sim 30$~km is well matched by planet formation models of runaway growth. Note the break in the size distribution at $R \sim 30$~km. We assumed an albedo of 4\% and a distance of 42~AU when converting the observed magnitudes into radii. We note here however, that the exact choice for the value of the albedo does not affect the fit between the observational data and the numerical results because assuming a different value for the albedo would simply shift the x-axis values by a constant and this shift can be matched by the numerical results by letting the self-similar growth continue to larger/smaller sizes.}
\label{fig1}
\end{figure}

\begin{figure} [htp]
\centerline{\epsfig{file=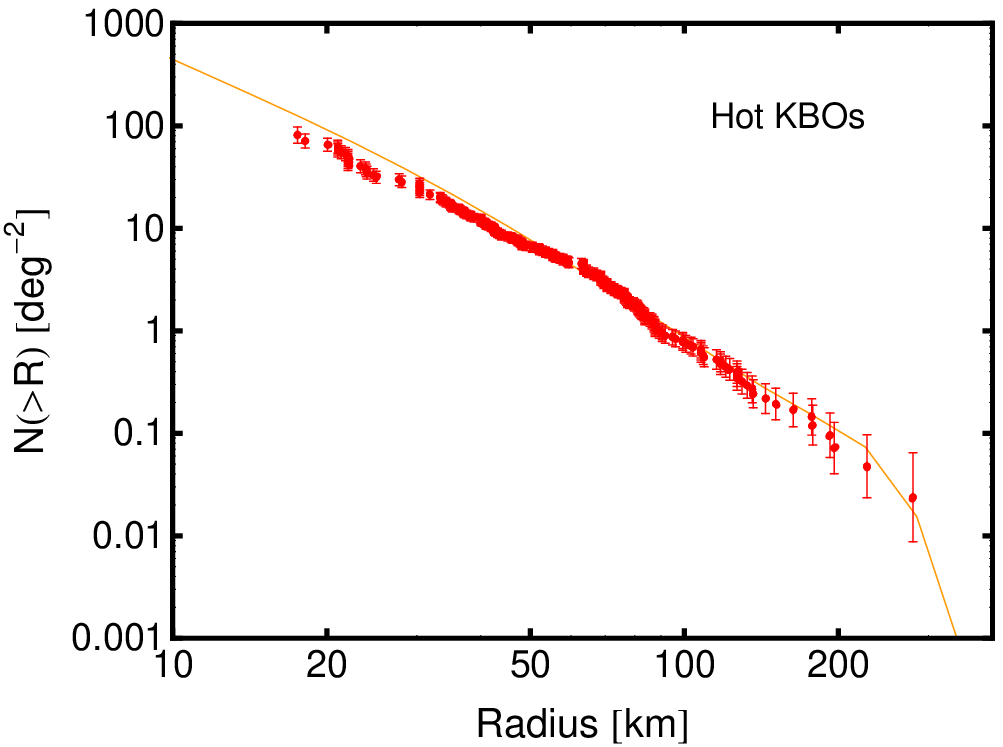, scale=1.2}}
\caption{Comparison between the observed Kuiper Belt size distribution for objects with inclinations $> 5^{\circ} $, also referred to as the hot population, as summarized in \citet{FH10} (points), with the numerical coagulation results from \citet{SS11} (line). The error bars give the $1 \sigma$ errors on the cumulative size distribution. The observed Kuiper Belt size distribution above $R \sim 30$~km is well matched by planet formation models of runaway growth. Note the break in the size distribution at $R \sim 30$~km does not seem as strongly pronounced as in Figure \ref{fig1}.}
\label{fig2}
\end{figure}

Figures \ref{fig1} and \ref{fig2} show a direct comparison between the results of runaway growth from the coagulation model from \citet{SS11} and the observed size distribution of dynamically cold and hot KBOs, respectively. The observed KBO size distribution was derived by \citet{FH10} by combining results from KBO surveys by \citet{CB99,GEL01,TJL01,AEL02,BTA04,PEL06,FK108,FH108,FH08,FK09}, and \citet{FH10}. Dynamically cold refers here to objects with inclinations less than $5^{\circ}$, whereas dynamically hot corresponds to those with $i>5^{\circ}$. The agreement of the observations with the simple coagulation model from \citet{SS11} is good. Figures \ref{fig1} and \ref{fig2} show that the cold and hot populations can both be fit by the same size distribution, with the notable difference that the largest bodies in each population grew to different typical sizes.

Provided that the fall off at large KBO sizes is not due to some selection effect, this suggests that KBOs grew to typical radii of about 100~km in the cold population and to typical radii of about 300~km in the hot population. If the growth in the hot and cold populations was terminated simultaneously, presumably by the excitation of the velocity dispersion of the growing KBOs and the smaller planetesimals, then this suggests that the hot population may have formed closer to the Sun than the cold population, because the shorter orbital periods and likely higher mass surface densities ensure faster growth at smaller semi-major axis. 


Observations, including the data plotted in Figures \ref{fig1} and \ref{fig2}, suggest that there is a break at around 30~km in the KBO size distribution  \citep[e.g.][]{BTA04,FH108,FK09,SO09,FH10}.  This break is usually attributed to collisional evolution of bodies with $R < 30$~km over the age of the solar system \citep[e.g.][]{D69,KB04,PS05}. The KBO size distribution below radii of $\sim 10$~km is still poorly constrained, because KBOs of these sizes are too small to be detected in reflected light. They can, however, be detected indirectly, by stellar occultations. Recent KBO occultation surveys provide the first estimates for the abundance and upper limits of km sized to sub-km sized KBOs \citep[e.g.][]{LC08,SO09,B10,SO12,ZL13}.

The work presented in this paper focuses on the size distribution of small KBOs below the break. We model the growth and the subsequent collisional evolution in the Kuiper Belt self-consistently by following the collisional evolution over 4.5 Gyrs of the whole KBO size distribution that resulted from runaway growth. We find that the break radius at $R\sim 30$~km and size distribution below the break are well matched by collisional evolution models that use published strength laws and make testable predictions for the small KBO size distribution. We show that the excess mass in small planetesimals from the runaway growth phase leaves a permanent signature in the size distribution of small bodies that is not erased after 4.5 Gyrs of collisional evolution. Observations of the small KBO size distribution can therefore test if large KBOs grew as a result of runaway growth and constrain the initial planetesimal sizes. 

This paper is structured as follows: We describe our Kuiper Belt growth and collisional evolution model in section 2. In section 3 we present our results and compare them with current observational constrains on small objects in the Kuiper Belt. Discussion and conclusions follow in section 4.

\section{Kuiper Belt Growth and Collisional Evolution Model}

\subsection{Growth Model}
We use the same coagulation model as described in \citet{SS11}, which follows the mass growth and the coupled evolution of the velocity dispersion using Safronov's statistical approach \citep{S72}. We refer the reader to \citet{SS11} for the full set of equations for the growth rates of the bodies in the different mass bins and for the corresponding evolution of their velocity dispersions. We investigate the KBO growth in a single annulus centered at 40~AU from the Sun with a width of 10~AU and start
the simulations with a total mass of about 20 Earth masses in small planetesimals. This mass surface density was derived by extrapolating of the minimum mass solar nebula \citep{H81} to 40~AU, after it has been enhanced by a factor of a few as required for the formation of Uranus and Neptune \citep[e.g.][]{GLS042,DB10}.  We assume that when the relative velocity exceeds the escape velocity of the larger of the two bodies (i.e., $v_{rel} > v_{esc_B}$), no accretion occurs and that, if the center of mass collisional energy of two colliding bodies exceeds the catastrophic destruction threshold, fragmentation takes place (see section \ref{s1} for details).

In the Kuiper Belt planet formation never went all the way to completion. The growth was likely terminated due to the excitation of the velocity dispersion of the growing KBOs and small planetesimals by the formation and migration of the planets in the outer solar system. We model this dynamical excitation by increasing the velocity dispersion of all bodies in our numerical model to 1~$\rm{km s^{-1}}$, which corresponds roughly to the random velocity dispersion in the Kuiper Belt today, once objects with the size of Pluto have formed. As long as most of the initial mass resides in planetesimals that are about 1~km in size or larger, destructive collisions and fragmentation are not important until objects comparable to the size of Pluto have formed. This is because initially the planetesimal velocities are smaller than their escape velocities and even as their velocity dispersions are stirred by the growing KBOs, objects of at least several hundreds of kilometers in radius have to form until they can dynamically excite the velocity dispersion of the small planetesimals above speeds needed for destructive collisions.\footnote{If initially most of the mass resided in planetesimals that were much smaller than 1~km in size, then the KBO growth maybe substantially different from the case investigated here, because destructive collisions and fragmentation will commence before bodies of a few hundred kilometers in radius have formed.} This picture changes completely once the velocity dispersions of all bodies are excited to 1~$\rm{km s^{-1}}$. From this time onward the growth is essentially terminated and destructive collisions lead to the onset of a collisional cascade. We assume, although objects in the hot and cold population likely formed at somewhat different locations, that they evolve together collisionally over the age of the solar system. This assumption is likely valid because the same physical processes, i.e., the formation and migration of Neptune, 
that are responsible for the excitations of the KBOs' velocity dispersion are responsible for placing the hot population into its current location. The KBO formation timescales are generally found to be less than 100~Myrs \citep[e.g.][]{KL99,SS11}, which suggests that the Kuiper Belt had close to 4.5 Gyrs to evolve collisionally. Observations of the Kuiper Belt size distribution find that the break radius and the slope of the size distribution below the break are the same in both the hot and cold KBO population, which is consistent with the idea that these two populations are undergoing collisional evolution together \citep{FH10}.

\subsection{Collisional Model}\label{s1}
We model destructive collisions in the following way. The catastrophic destruction threshold, $Q^*_D$, is defined as the specific energy needed to disperse the targets into a spectrum of individual objects such that the largest one has exactly half the mass of the original target. When the center of mass collisional energy of two colliding bodies, $m_1$ and $m_2$, exceeds the catastrophic destruction threshold, $Q^*_D$, then the combined mass, $m_1+m_2$, is distributed such that one body of mass $0.5 (m_1+m_2)$ is formed and the remaining mass is distributed as debris over all mass bins that correspond to planetesimal sizes with $m < 0.5 (m_1+m_2)$ according to a differential power law size distribution given by $dN/dR \propto R^{-q^*}$. 

\begin{figure} [htp]
\centerline{\epsfig{file=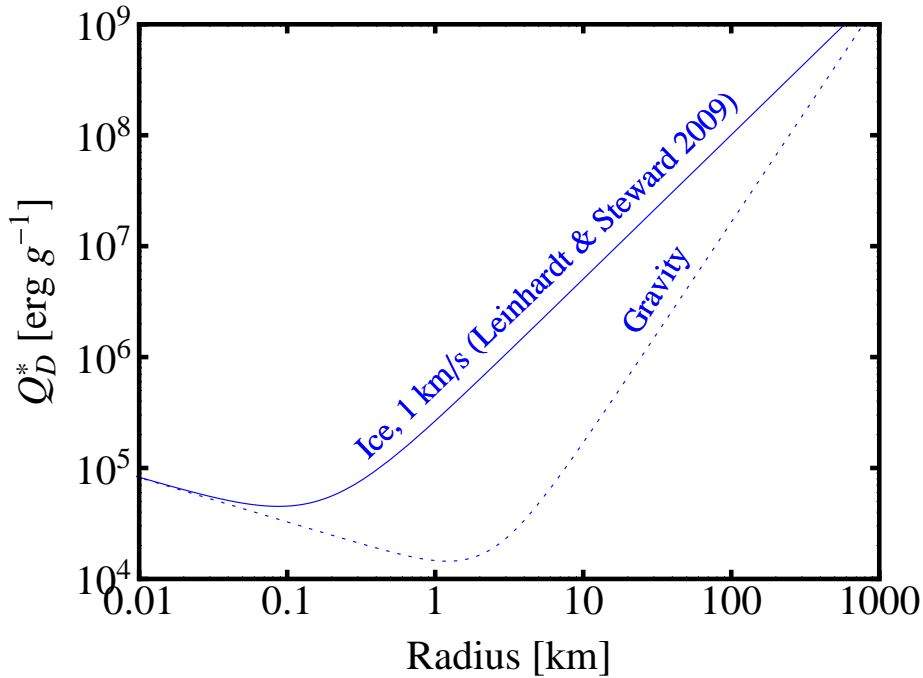, scale=1.2}}
\caption{Catastrophic destruction threshold, $Q^*_D$, as a function of size. The solid, blue line corresponds to results from \citet{LS09} for ice and $1~\rm{km~s^{-1}}$ impact velocities, which corresponds to the velocity dispersion in the Kuiper Belt today. For comparison, the catastrophic destruction threshold corresponding to the  gravitational binding  energy in the gravity regime and the same material strength law as before is shown as dashed blue line. The gravitational binding energy gives an absolute lower limit to the catastrophic destruction threshold, since bodies cannot be weaker than this.}
\label{fig8}
\end{figure}

Since the Kuiper Belt consists of mostly icy bodies with an average velocity dispersion of about $1~\rm{km~s^{-1}}$, we adopt the strength law from \citet{LS09} for ice and 1~$\rm{km~s^{-1}}$ impact speeds for the catastrophic destruction threshold, which is given by
\begin{equation}\label{e676}
Q^*_D=1.3 \times 10^6 \left(\frac{R}{1~\rm{cm}}\right)^{-0.4}+0.08 \left(\frac{R}{1~\rm{cm}}\right)^{1.3} \rm{erg~g^{-1}}.
\end{equation}
Figure \ref{fig8} shows $Q^*_D$ as a function of size and the transition from the gravity dominated regime ($R \gtrsim 0.1$~km) to the material strength dominated regime ($R \lesssim 0.1$~km). For comparison, the catastrophic destruction threshold corresponding to the specific gravitational binding  energy  in the gravity regime and the same material strength law as before is also shown in Figure \ref{fig8}. The gravitational binding energy gives an interesting absolute lower limit to the catastrophic destruction threshold, since bodies cannot be weaker than this.

For the fragment size distribution, $dN/dR \propto R^{-q^*}$, we adopt $q^*=3.68$. This value of $q^*$ corresponds to the expected collisional equilibrium size distribution, which has a power law index that is given by
\begin{equation}\label{e1}
q_{eq}=\frac{21+\alpha}{6+\alpha}
\end{equation}
where $\alpha$ is the exponent of $R$ in $Q^*_D$ (see Equation (\ref{e676}) in the material strength dominated regime \citep[e.g.][]{PS12}. From Equation (\ref{e1}) we find that $\alpha=-0.4$ yields $q_{\rm{eq}} = 3.68$.

\section{Results}
Combining our growth and collisional model we investigated the evolution of the KBO size distribution starting from various initial planetesimal sizes over 4.5 Gyrs.

\subsection{1~km sized Planetesimals}
\begin{figure} [htp]
\centerline{\epsfig{file=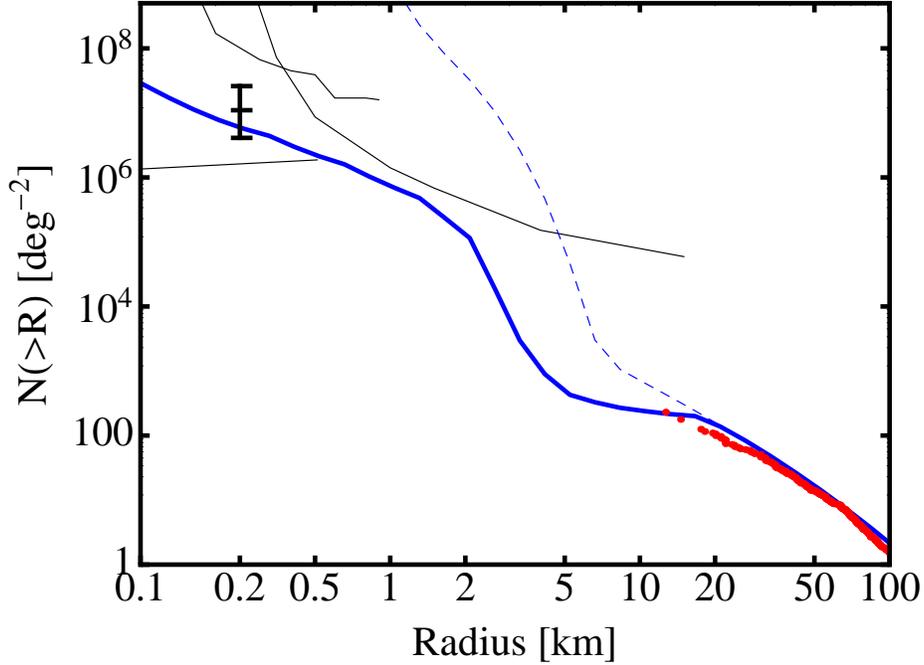, scale=1.2}}
\caption{The small KBO size distribution after 4.5 Gyrs of collisional evolution for an initial planetesimal population that consisted of 1~km radii objects (blue thick line). For comparison, the KBO size distribution at the end of runaway growth and at the onset of destructive collisions is given by the dashed blue line. The observed KBO size distribution is shown by the red points \citep{FH10}. The black point with error bars and the thin black lines ranging from 0.1~km to 1~km represent the best estimate and the 95\% upper and lower limits on the small KBO population from the HST-FGS occultation survey by \citet{SO12}, respectively. The thin black line ranging from 0.2~km to 20~km represents the 95\% upper limit on the small KBO population from TAOS \citep{ZL13}.}
\label{fig7}
\end{figure}

\begin{figure} [htp]
\centerline{\epsfig{file=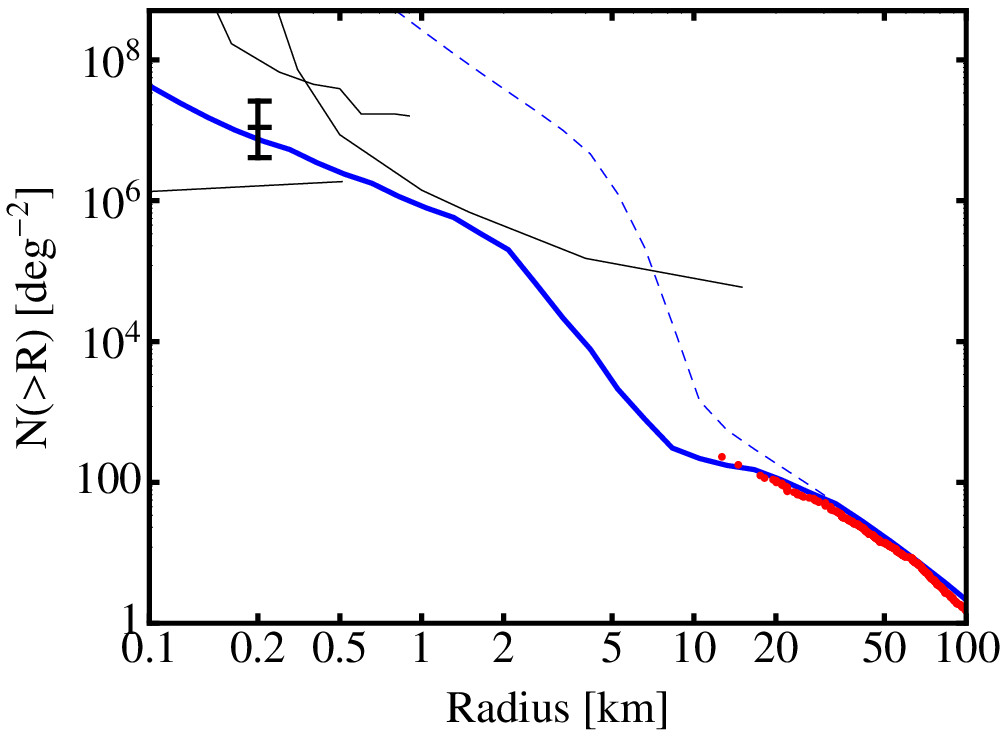, scale=1.2}}
\caption{The small KBO size distribution after 4.5 Gyrs of collisional evolution for an initial planetesimal population  that contained equal mass per logarithmic mass bin in bodies ranging from 0.4~km to 4~km in radius (blue thick line). For comparison, the KBO size distribution at the end of runaway growth and at the onset of destructive collisions is given by the dashed blue line. The observed KBO size distribution is shown by the red points \citep{FH10}. The black point with error bars and the thin black lines ranging from 0.1~km to 1~km represent the best estimate and the 95\% upper and lower limits on the small KBO population from the HST-FGS occultation survey by \citet{SO12}, respectively. The thin black line ranging from 0.2~km to 20~km represents the 95\% upper limit on the small KBO population from TAOS \citep{ZL13}.}
\label{fig3}
\end{figure}

Figures \ref{fig7} and \ref{fig3} show the resulting KBO size distribution (solid blue line) after 4.5 Gyrs of growth and collisional evolution when starting from an initial planetesimal size distribution that consists solely of 1~km sized bodies, and from an initial planetesimal size distribution that has equal mass per logarithmic mass bin for bodies ranging from 0.4~km to 4~km in radius, respectively. For comparison, the dashed blue lines in Figures \ref{fig7} and \ref{fig3} show the KBO size distribution at the end of runaway growth just before the start of destructive collisions. First of all it is interesting to note that the resulting small KBO size distributions do not follow a single power law below the break (i.e., below $R \sim 30$~km) as one may naively expect. Instead we find that the small KBO size distribution exhibits a strong deficit of bodies around $R \sim 10$~km in size and a strong excess of bodies around 2~km in radius compared to abundances from a single power law size distribution spanning the range from 0.1~km to 30~km. This deficit and excess are caused by the planetesimal size distribution left over from the runaway growth phase, which left most of the initial mass in small planetesimals. This excess mass in small planetesimals leaves a permanent signature in the size distribution of small bodies that is not erased after 4.5 Gyrs of collisional evolution. The resulting KBO size distributions shown in Figures \ref{fig7} and \ref{fig3} are both consistent with abundance estimates and upper limits from KBO occultation surveys \citep{SO12,ZL13} shown in black. However, if all the mass resides solely in 1~km planetesimals initially, not quite enough mass is depleted in 10-30~km radius range compared to observations (see Figure \ref{fig7}). If on the other hand, we start with an initial planetesimal size distribution that has equal mass per logarithmic mass bin for bodies ranging from 0.4~km to 4~km in radius we find good agreement with the observations (see Figure \ref{fig3}).

\begin{figure} [htp]
\centerline{\epsfig{file=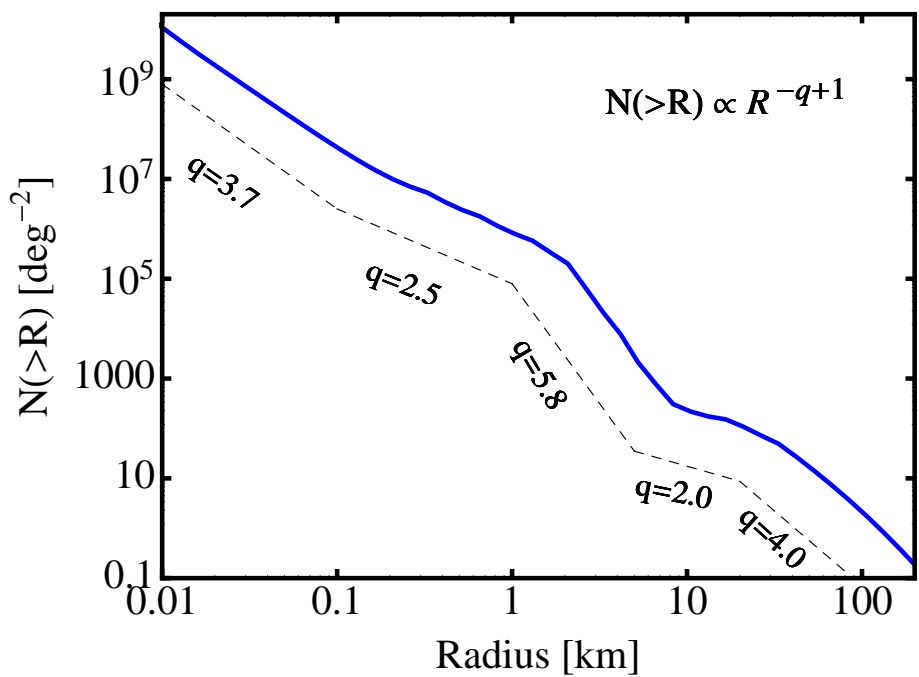, scale=1.2}}
\caption{The same small KBO size distribution after 4.5 Gyrs of collisional evolution as shown in Figure \ref{fig3} but plotted with the corresponding differential power law indices for the different segments of the size distribution. The deficit around 10~km results from an excess of  $\sim 1$~km planetesimals at the onset of the collisional evolution. The size distribution for $R \lesssim 0.1$~km takes on the expected equilibrium value for material strength dominated bodies as calculated in section 2.2. The size distribution above $R\sim 30$~km remains unchanged by collisional evolution over the age of the solar system and is therefore primordial.}
\label{fig5}
\end{figure}

\begin{figure} [htp]
\centerline{\epsfig{file=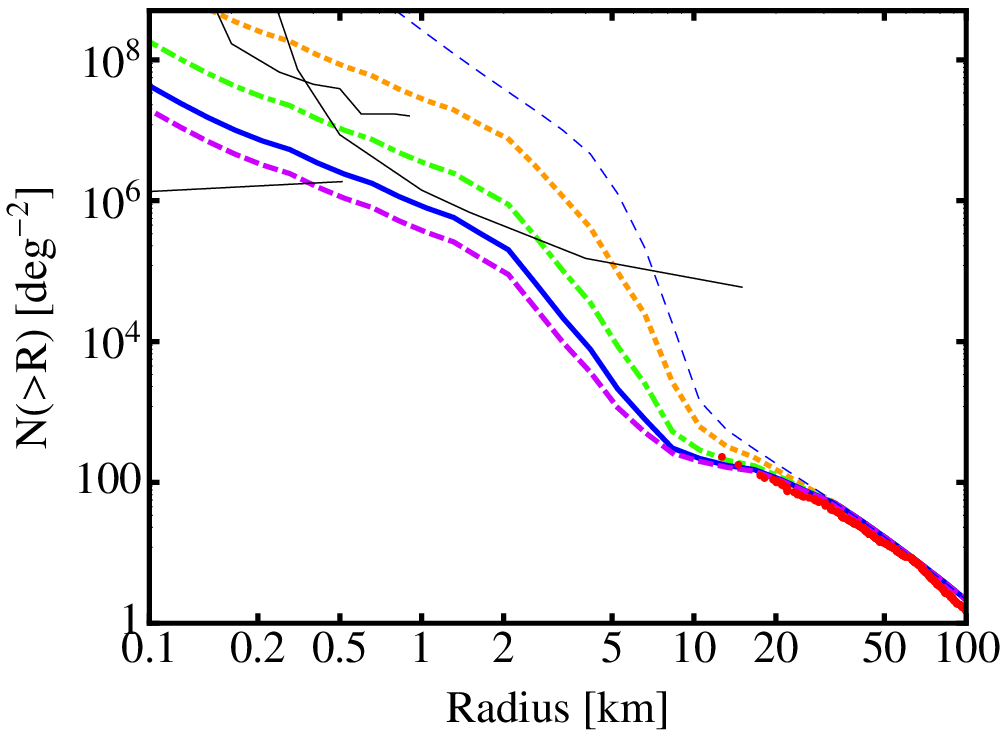, scale=1.1}}
\caption{The time evolution of the small KBO size distribution for an initial planetesimal population  that contained equal mass per logarithmic mass bin in bodies ranging from 0.4~km to 4~km in radius. The thin blue line corresponds to the KBO size distribution at the end of runaway growth before the onset of destructive collisions; the yellow, dotted line corresponds to 100 Myrs; the green dot-dashed line to 1 Gyr; the solid, blue line to 4.5 Gyrs, and the dashed purple line to 10 Gyrs of collisional evolution. The observed KBO size distribution is shown by the red points \citep{FH10}. The thin black lines ranging from 0.1~km to 1~km represent the 95\% upper and lower limits on the small KBO population from the HST-FGS occultation survey by \citet{SO12} and the thin black line ranging from 0.2~km to 20~km represents the 95\% upper limit on the small KBO population from TAOS \citep{ZL13}.}
\label{fig9}
\end{figure}

Figure \ref{fig5} shows the same collisionally evolved size distribution as in Figure \ref{fig3} but with the corresponding power law indices for the different segments. KBOs with $R \gtrsim 30$~km follow a size distribution with a differential power law index $q \sim 4$, which is a relic from their formation and has not been modified by collisional evolution over 4.5 Gyrs. The power law index of the size distribution between 0.1~km to 30~km changes from $q \sim 2$ ($10~\rm{km} \lesssim R\lesssim 30$~km) to $q \sim 5.8$ ($2~\rm{km} \lesssim R\lesssim 10$~km) and then to $q \sim 2.5$ ($0.1~\rm{km} \lesssim R\lesssim 2$~km).  This change in the slopes of the size distributions is mainly caused by the excess population of planetesimals that was left over from the runaway growth phase. This excess population gives rise to a very steep size distribution between $\sim 2$~km and $\sim 10$~km which grows shallower in time (see Figure \ref{fig9}) because the excess in kilometer-sized planetesimals is being depleted with time. The shallow, $q=2.0$, power law index between $\sim 10$~km and $\sim 30$~km is due to the excess population of kilometer-sized planetesimals that started to deplete the population of bodies between $\sim 10$~km to $\sim 30$~km. The size distribution for $R \lesssim 0.1$~km takes on the expected equilibrium value for material strength dominated bodies as calculated in section 2.2 from Equation (\ref{e1}) for the catastrophic destruction threshold from \citet{LS09} for ice and impact velocities of $1~\rm{km~s^{-1}}$. The precise values of the power law index in the different size regimes and the exact location of the inflection points depend on the  catastrophic destruction criterion as a function of radius and the initial planetesimal size distribution. For example, starting with planetesimal sizes that range from 0.4~km to 4~km with equal mass per logarithmic mass interval yields a smaller value for $q$ in the 1 to 10~km range than starting with all the mass in 1~km-sized planetesimals (see Figures \ref{fig7} and Figures \ref{fig3} for comparison).  
However, the overall behaviour, i.e., a deficit of bodies around $R\sim 10$~km and an excess of bodies around 2~km in radius, does not depend on the exact choices of the catastrophic destruction criterion and the initial planetesimal size distribution. 

Figure \ref{fig9} displays the time evolution of the small KBO size distribution. The differential power law indices between $\sim10$~km and $\sim 30$~km and between $\sim 2$~km and $\sim 10$~km become shallower with time. The decrease in the power law index between $\sim 2$~km and $\sim 10$~km is due to the fact that the excess population of planetesimals that was left over from the runaway growth, which gave rise to a very steep size distribution between $\sim 2$~km and $\sim 10$~km (dashed blue line in Figure \ref{fig9}), is being depleted by destructive collisions with time. The decrease in the power law index between $\sim 10$~km and $\sim 30$~km is due to the excess population of kilometer-sized planetesimals that start to deplete the population of bodies between $\sim 10$~km to $\sim 30$~km. I.e., the catastrophic destruction threshold from \citet{LS09} yields, for a velocity dispersion of $1~\rm{km~s^{-1}}$, that 10~km sized bodies are typically destroyed by bodies $\sim 1$~km in radius and 30~km sized bodies are typically destroyed by bodies that are $\sim 8$~km in radius. The power-law index below $\sim 2$~km evolves to $q \sim 2.5$ and remains close to constant from then onwards.

\subsection{10~km sized Planetesimals}
\begin{figure} [htp]
\centerline{\epsfig{file=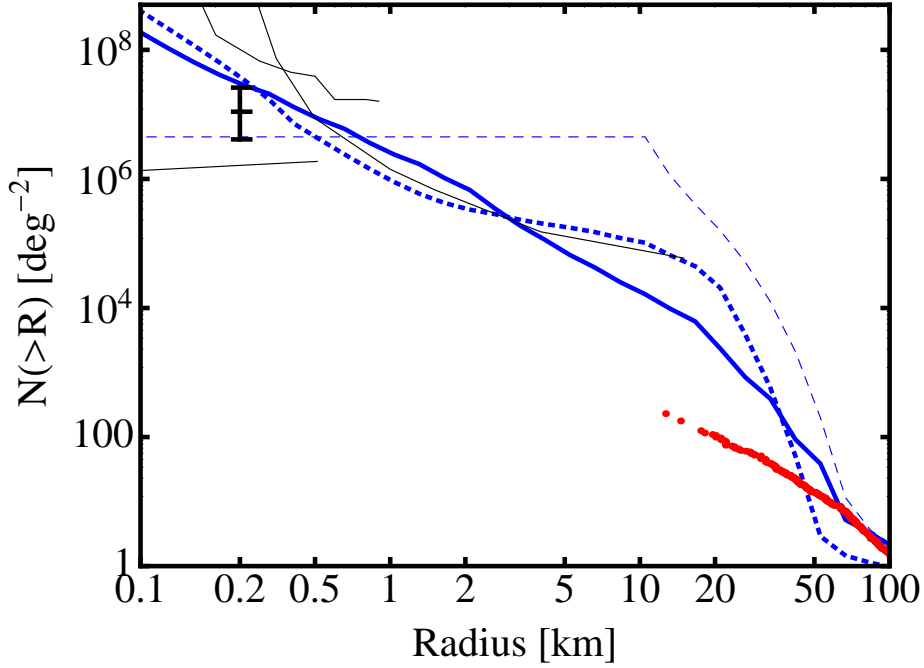, scale=1.2}}
\caption{The small KBO size distribution after 4.5 Gyrs of collisional evolution for an initial planetesimal population that consisted of 10~km radii bodies (blue thick line). The resulting size distribution assuming that large KBOs are solely held together by their own gravity is shown by the dotted blue line. For comparison, the KBO size distribution at the end of runaway growth and at the onset of destructive collisions is given by the dashed blue line. The observed KBO size distribution is shown by the red points. The black point with error bars and the thin black lines ranging from 0.1~km to 1~km represent the best estimate and the 95\% upper and lower limits on the small KBO population from the HST-FGS occultation survey by \citet{SO12}, respectively. The thin black line ranging from 0.2~km to 20~km represent the 95\% upper limit on the small KBO population from TAOS \citep{ZL13}. The resulting KBO size distributions are inconsistent with the observed KBO size distribution for bodies with radii ranging from 10~km to 100~km and with upper limits from the TAOS KBO occultation survey \citep{ZL13}. This results holds true even if we assume that large KBOs are only held together by their own gravity. We conclude that the Kuiper Belt did not form from 10-km sized planetesimals by coagulation, because not enough of the initial planetesimals can be ground down over the age of the solar system to match observations.}
\label{Fig6}
\end{figure}

Figure \ref{Fig6} shows the resulting small KBO size distribution after 4.5 Gyrs of collisional evolution for an initial planetesimal population that consisted of 10~km radii bodies (solid blue line). The small KBO size distribution is inconsistent with the observed size distribution of KBOs with radii ranging from 10~km to 100~km (red points) and with upper limits from the TAOS KBO occultation survey \citep{ZL13}. Even if we assume that large KBOs are only held together by their own gravity (blue dotted line in Figure \ref{Fig6}), which is an absolute lower limit on their strength, because bodies can't be weaker than this, we find that we cannot match the observed KBO size distribution. We also started with initial planetesimal populations that contained equal mass per logarithmic mass interval between 1~km and 10~km in radius and were still unable to find a reasonable agreement between the resulting small KBO size distribution and the observations. This leads us to conclude that the Kuiper Belt did not form via coagulation from an initial planetesimal population that contained most of the initial mass in planetesimals that were 10~km in radius or larger, because not enough of the initial planetesimals can be ground down over the age of the solar system to match observations.

\section{DISCUSSION AND CONCLUSIONS}
We studied the size distribution of small KBOs by modeling self-consistently the growth and the subsequent collisional evolution over 4.5 Gyrs in the Kuiper Belt and arrive at the following results:

$\bullet$ The Kuiper Belt size distributions of the cold and hot population for radii $\gtrsim 30$~km are primordial and can both be well fit by the resulting size distributions from planet formation models of runway growth (see Figure \ref{fig1} and \ref{fig2}) with the notable difference that the largest bodies in the hot population grew to larger radii than in the cold population, which is consistent with the idea that the hot population formed at smaller semi-major axis compared to the cold population.

$\bullet$ The break radius at $R\sim 30$~km and size distribution below the break are well matched by collisional evolution models that use published strength laws \citep{LS09} and start with resulting size distributions from runway growth and follow the collisional evolution in the Kuiper Belt over 4.5 Gyrs.

$\bullet$ Compared to a single power law size distribution that would span the whole range from 0.01~km to 30~km, we find in general a strong deficit of bodies around $R\sim 10$~km and a strong excess of bodies around 2~km in radius. This deficit and excess are caused by the planetesimal size distribution left over from the runaway growth phase, which leaves most of the initial mass in small bodies. This excess mass in small planetesimals leaves a permanent signature in the size distribution of small bodies that is not erased after 4.5 Gyrs of collisional evolution. Future KBO occultation surveys, which probe the small KBO size distribution, can therefore test if large KBOs grew as a result of runaway growth and constrain the initial planetesimal sizes.

$\bullet$ The observed KBO size distribution derived by \citet{FH10} by combining various KBO surveys \citep{CB99,GEL01,TJL01,AEL02,BTA04,PEL06,FK108,FH108,FH08,FK09,FH10} and results from recent optical KBO occultation surveys \citep{SO12,ZL13} are best matched by an initial planetesimal population that contained about equal mass in bodies ranging from 0.4~km to 4~km in radius. In addition, the resulting KBO size distribution after 4.5 Gyrs of collisional evolution is also consistent with upper limits from KBO occultation surveys at x-ray wavelengths that probe objects ranging from $\sim 30$~m to 300~m in radius \citep{JLM08,CCC11,CL12}.

$\bullet$ The observed KBO size distribution for $R > 10$~km cannot be matched if most of the initial planetesimal mass resided in bodies that were 10~km in radius or larger, because their resulting size distribution cannot be sufficiently depleted over 4.5 Gyrs to match observations. We conclude from this that the Kuiper Belt did not form by coagulation from an initial planetesimal size distribution that contained most of its mass in planetesimals with radii of 10~km or larger.

There are several further interesting things to note here: 

Since the excess mass in small planetesimals from the runaway growth phase leaves a permanent signature in the size distribution of small bodies that is not erased after 4.5 Gyrs of collisional evolution, future KBO occultation surveys will be able to test whether large KBOs grew as a result of runaway growth from an initial planetesimal population consisting of bodies ranging from a few hundred meters to a few kilometers in size. The small KBO size distribution therefore offers the opportunity to observationally constrain the initial planetesimal sizes from which planets form, which remains one of the major open questions in planet formation theory \citep{CY10}.

The resulting small KBO size distributions that we find all contain enough bodies to satisfy the required supply rate for the Jupiter family comets \citep{VM08}. If the Kuiper belt formed by coagulation from km-sized planetesimals then there should be an excess of about a factor of 1000 of small comets with initial radii (i.e., before any mass loss or break up occurs) of 2~km compared to 10~km.

Because the comet size distribution has likely been modified by mass loss and break up of the cometary nuclei, the size distribution of centaurs should provide a more reliable probe of the KBO size distribution between 1 to 10~km in radius. Unfortunately, although about two hundred centaurs are currently known with sizes ranging from about 100~km to 1~km, no well characterized survey has been carried out  to date that would allow the derivation of a de-biased centaur size distribution and therefore probe the small KBO size distribution.

Finally, it is very intriguing that there is a striking similarity between the small KBO size distribution that we find after 4.5 Gys of collisional evolution and the reconstructed impact size distribution from the cratering records on the Saturnian satellites \citep{MR12}. \citet{MR12} find that the cratering size distribution of the old terrains of Dione, Hyperion, Iapetus, Mimas, Phobe, Rhea and Tethys can be explained by a single impactor population that follows a size distribution with differential power laws indices of $q=4$ for $R> 30$~km, $q=2.0$ for $10~\rm{km} < R < 30$~km, $q=4.2$ for $1~\rm{km}<R<10$~km, $q=2.6$ for $0.1~\rm{km} < R < 1$~km and $q=3.7$ for $0.01~\rm{km} < R < 0.1$~km. These values are in remarkably good agreement with the power laws indices that we find for the small KBO size distribution and that are shown in Figures \ref{fig5}. The only notable difference between our small KBO size distribution and the results from \citet{MR12} seems to be in the range from $\sim 1$~km to $\sim 10$~km for which we find a steeper size distribution with power law index $q=5.8$.
The similarities between our results for the small KBO size distribution and the reconstructed impactor size distribution suggest that the impactors that bombarded the Saturnian moons originated from the Kuiper Belt.

\acknowledgements{We thank David Jewitt for his comments and suggestions.}

\bibliographystyle{aj} 

\end{document}